\documentclass[aps,prl,superscriptaddress,twocolumn]{revtex4}
\usepackage{amsmath}
\usepackage{amssymb}
\usepackage{graphicx}
\usepackage{dcolumn}
\usepackage{bm}

\begin{document}

\title{Imaginary Potential Induced Quantum Coherence for Bose-Einstein
Condensates}
\author{Hongwei Xiong}
\email{xionghongwei@wipm.ac.cn} \affiliation{State Key Laboratory of
Magnetic Resonance and Atomic and Molecular Physics, Wuhan Institute
of Physics and Mathematics, Chinese Academy of Sciences, Wuhan
430071, P. R. China}
\date{\today }

\begin{abstract}
The role of complex potentials in single-body Schr\H{o}dinger equation has
been studied intensively. We study the quantum coherence for degenerate Bose
gases in complex potentials, when the exchange symmetry of identical bosons
is considered. For initially independent Bose-Einstein condensates, it is
shown that even very weak imaginary potential can induce perfect quantum
coherence between different condensates. The scheme to observe imaginary
potential induced quantum coherence is discussed.

\end{abstract}

\maketitle

The extension of quantum mechanics to complex potentials has been studied
intensively in diverse areas of physics \cite{Benderreview,Muga}. The
Hamiltonian of the system with complex potentials is not Hermitian any more.
How to understand and reveal the fundamental properties of non-Hermitian
Hamiltonians is very interesting, especially after Bender \textit{et al. }%
\cite{Bender} found the counter-intuitive result that non-Hermitian
Hamiltonians have entirely real eigenvalue spectra for some physical systems
satisfying the parity-time symmetry. Besides an alternative formulation of
quantum mechanics \cite{Benderreview} with complex potentials, it has
potential applications in a lot of different systems, such as molecular
collisions \cite{Muga}, matter wave dynamics \cite%
{Oberthaler,Keller,Oberthaler1,Bernet,Stuzle}, light propagation \cite%
{Makris,Guo,Longhi}, optical solitons \cite{Muss}, and quantum transport
\cite{Dogan} \textit{etc.}.

Ultracold atomic gases are promising to study quantum behavior with complex
potentials, because various complex potentials can be realized and
manipulated experimentally \cite{Oberthaler,Keller,Oberthaler1,Bernet,Stuzle}%
. Although all experiments up to date \cite%
{Oberthaler,Keller,Oberthaler1,Bernet,Stuzle} are not about degenerate
gases, the remarkable advances of degenerate Bose and Fermi gases \cite%
{BECreview} make it feasible to study quantum many-body physics with complex
potentials. Quite surprising, to our best knowledge, both theoretical and
experimental studies on quantum many-body physics in complex potentials are
highly scarce, although the single-body problem has been intensively studied
for cold atoms \cite%
{Oberthaler,Keller,Oberthaler1,Bernet,Stuzle,Berry,Berry1}. An important
reason for this situation lies in the common belief that the external
potential itself will not influence quantum many-body behavior, without
considering interatomic interaction. Different from this common belief, our
work shows that complex potentials may play an important role in quantum
many-body behavior. Based on the many-body Schr\H{o}dinger equation for
identical bosons, we focus our studies on the quantum coherence for two
initially independent Bose-Einstein condensates (BECs) in a complex periodic
potential. It is found that even a very weak imaginary potential can induce
perfect quantum coherence between two condensates. Our theoretical work
gives an example that complex potentials not only play a role in single-body
problem, but also play important role in quantum many-body behavior. This
result opens the way to study novel quantum many-body physics for ultracold
gases \cite{Bloch} with complex potentials.

Complex potentials can be generated with the interaction between near
resonant laser and a two-level atom with an additional decay channel of the
excited state to another state \cite%
{Oberthaler,Keller,Oberthaler1,Bernet,Stuzle}. The complex potential takes
the following general form \cite{Chus}%
\begin{equation}
V\left( \mathbf{r}\right) =\frac{d_{e}^{2}\mathbf{E}^{2}\left( \mathbf{r}%
\right) }{\hbar \left( \delta +i\Gamma /2\right) }.  \label{complexpotential}
\end{equation}%
Here, $d_{e}$ is the dipole matrix element of an atom, while $\delta $ is
the detuning of the light frequency. $\Gamma $ is the loss rate of the
excited level through the additional decay channel. $\mathbf{E}\left(
\mathbf{r}\right) $ is the electric field of the laser. By varying the
detuning, one can vary the potential from a perfect real potential to an
imaginary potential. The manipulation of the electric field makes various
complex potentials possible such as complex periodic potential.

It has been shown that the solution of the single-particle Schr\H{o}dinger
equation with complex potentials given by (\ref{complexpotential}) agrees
well with the experimental results about cold atoms \cite%
{Oberthaler,Keller,Oberthaler1,Bernet,Stuzle}. Here we consider the solution
of the many-body Schr\H{o}dinger equation when both complex potentials and
exchange symmetry of identical bosons are considered. To show clearly the
imaginary potential induced quantum coherence between initially independent
condensates, we consider a process shown in Fig. 1. In Fig. 1a, there are
two initially independent condensates described by a Fock state $\left\vert
N_{1},N_{2}\right\rangle $, with $N_{1}$ and $N_{2}$ being the initial
particle numbers in two condensates. Initially, there is no overlapping
between these two condensates. As shown in Fig. 1b, these two condensates
are then allowed to expand after switching on a complex periodic potential.

\begin{figure}[tbp]
\includegraphics[width=0.75\linewidth,angle=270]{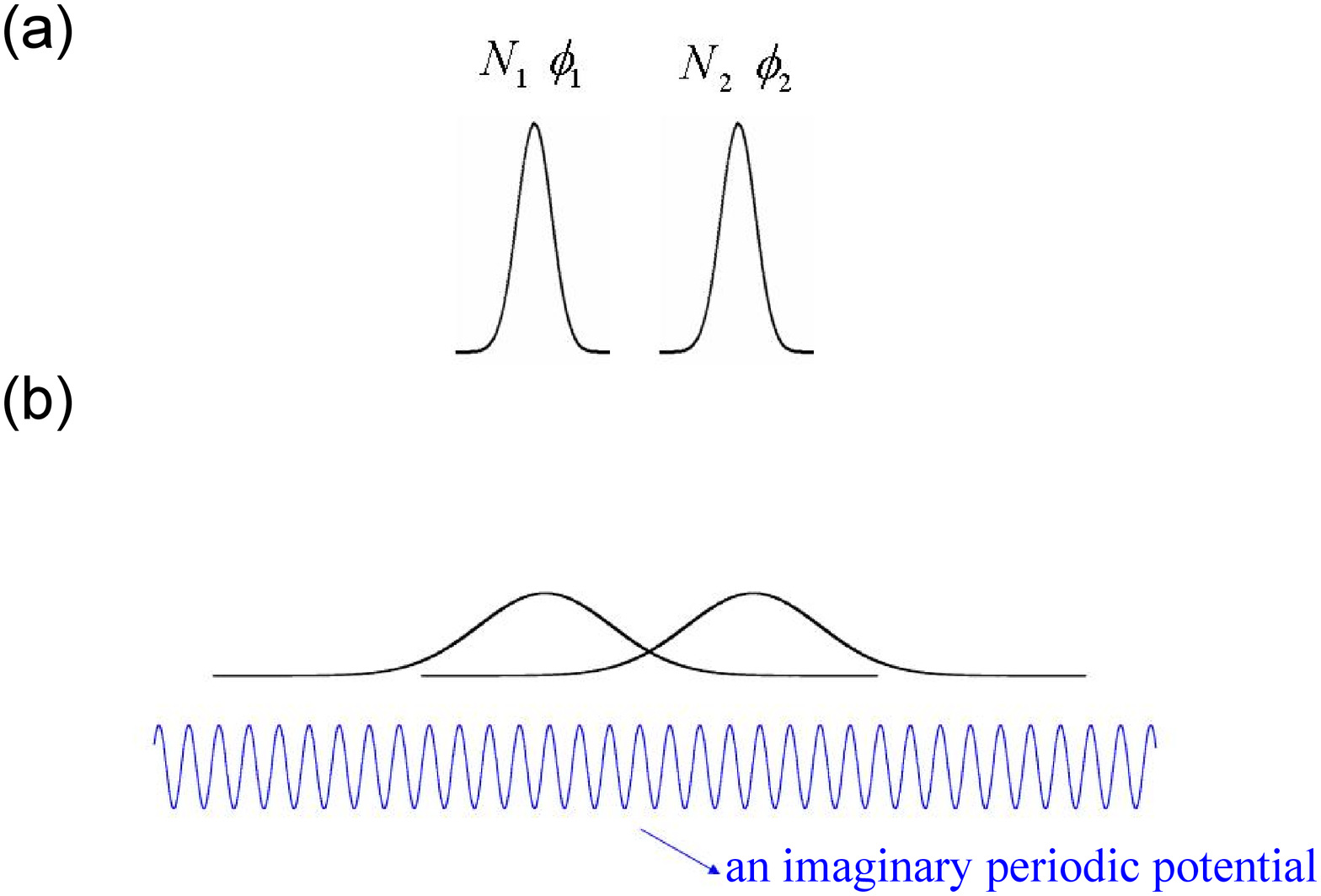}
\caption{In Fig. a, there are two initially independent Bose-Einstein
condensates. In Fig. b, a complex periodic potential is switched on after
these two condensates are allowed to expand.}
\end{figure}

When the exchange symmetry of identical bosons is considered, the many-body
wave function for identical bosons occupying two modes can be written as%
\begin{eqnarray}
&&\Psi \left( \mathbf{r}_{1},\mathbf{r}_{2},\cdots ,\mathbf{r}_{N},t\right)
=A_{n}\sqrt{\frac{N_{1}!N_{2}!}{\left( N_{1}+N_{2}\right) !}}  \notag \\
&&\sum\limits_{P}P\left[ \phi _{1}\left( \mathbf{r}_{1},t\right) \cdots \phi
_{1}\left( \mathbf{r}_{N_{1}},t\right) \times \right.  \notag \\
&&\left. \phi _{2}\left( \mathbf{r}_{N_{1}+1},t\right) \cdots \phi
_{2}\left( \mathbf{r}_{N_{1}+N_{2}},t\right) \right] ,  \label{wave-function}
\end{eqnarray}%
where $P$ denotes $\left( N_{1}+N_{2}\right) !/N_{1}!N_{2}!$ permutations
for the bosons in different single-particle wave functions $\phi _{1}$ and $%
\phi _{2}$. To give a general study, we assume $\zeta \left( t\right) =\int
\phi _{1}\left( \mathbf{r},t\right) \phi _{2}^{\ast }\left( \mathbf{r}%
,t\right) d\mathbf{r}$ from the beginning to consider the possible
nonorthogonality between $\phi _{1}$ and $\phi _{2}$. We assume further $%
\eta _{1}\left( t\right) =\int \left\vert \phi _{1}\left( \mathbf{r}%
,t\right) \right\vert ^{2}d\mathbf{r}$ and $\eta _{2}\left( t\right) =\int
\left\vert \phi _{2}\left( \mathbf{r},t\right) \right\vert ^{2}d\mathbf{r}$
to consider the decay of $\eta _{1}$ and $\eta _{2}$ due to complex
potentials. $A_{n}$ is introduced so that the average overall particle
number is $N_{1}\eta _{1}+N_{2}\eta _{2}$. After lengthy calculations, we
have
\begin{equation}
\left\vert A_{n}\right\vert ^{2}=\sum\limits_{i=0}^{\min \left(
N_{1},N_{2}\right) }\frac{N_{1}!N_{2}!\eta _{1}^{N_{1}-i}\eta
_{2}^{N_{2}-i}\left\vert \zeta \left( t\right) \right\vert ^{2i}\left(
N_{1}+N_{2}\right) }{i!i!\left( N_{1}-i\right) !\left( N_{2}-i\right)
!\left( N_{1}\eta _{1}+N_{2}\eta _{2}\right) }.
\end{equation}%
In this paper, to give a concise expression for various coefficients such as
$A_{n}$, we have introduced the rule $0^{0}=1$.

The many-body Schr\H{o}dinger equation for $\Psi $ is%
\begin{eqnarray}
i\hbar \frac{\partial \Psi }{\partial t} &=&\left[ \sum_{i=1}^{N}\left( -%
\frac{\hbar ^{2}}{2m}\nabla _{i}^{2}+V\left( \mathbf{r}_{i}\right) \right)
\right.  \notag \\
&&\left. +g\sum_{i<j}^{N}\delta \left( \mathbf{r}_{i}-\mathbf{r}_{j}\right)
-\mu \right] \Psi .  \label{many}
\end{eqnarray}%
Here $\mu $ is introduced due to the decay of $\eta _{1}$ and $\eta _{2}$.
The introduction of $\mu $ is a direct generalization of the evolution
equation for a single condensate with decay \cite{Ueda}. $g$ represents the
coupling due to interatomic collisions. To show clearly the role of complex
potentials, we consider the situation of $g=0$. The role of interatomic
collisions will be discussed at the end of this paper. Even the
nonorthogonality between $\phi _{1}$ and $\phi _{2}$ is considered, it is
not difficult to prove rigorously that the solution of $\Psi $ is given by%
\begin{eqnarray}
i\hbar \frac{\partial \phi _{1}}{\partial t} &=&\left( -\frac{\hbar ^{2}}{2m}%
\nabla ^{2}+V\right) \phi _{1},  \notag \\
i\hbar \frac{\partial \phi _{2}}{\partial t} &=&\left( -\frac{\hbar ^{2}}{2m}%
\nabla ^{2}+V\right) \phi _{2},  \notag \\
i\hbar \frac{\partial A_{n}}{\partial t} &=&-\mu A_{n}.  \label{phi1phi2}
\end{eqnarray}%
We stress that the expression of $\Psi $ given by Eq. (\ref{wave-function})
will not change in the whole dynamical process if $\phi _{1}$ and $\phi _{2}$
satisfy the above equation.

From Eq. (\ref{phi1phi2}), we have%
\begin{equation}
i\hbar \frac{d\zeta }{dt}=\int \left( V-V^{\ast }\right) \phi _{1}\phi
_{2}^{\ast }d\mathbf{r}.  \label{zeta}
\end{equation}%
As expected, $d\zeta /dt=0$ if $V$ is a real potential. However, if $V$ has
an imaginary component, $d\zeta /dt$ may be nonzero. This means that with
time evolution, $\zeta $ may be nonzero even for two initially orthogonal
wave functions. This is the reason why we give a general consideration of $%
\zeta $ from the beginning. In this situation, the exact expression of the
density expectation value is given by \cite{Xiong}:
\begin{equation}
n=a\left\vert \phi _{1}\right\vert ^{2}+2b\times \mathrm{Re}\left[
e^{i\varphi _{c}}\phi _{1}^{\ast }\phi _{2}\right] +c\left\vert \phi
_{2}\right\vert ^{2},  \label{density}
\end{equation}%
where the coefficients are%
\begin{eqnarray*}
a &=&\sum\limits_{i=0}^{\min \left( N_{1}-1,N_{2}\right) }\frac{%
A_{n}^{2}N_{1}!N_{2}!\left\vert \zeta \left( t\right) \right\vert ^{2i}}{%
i!i!\left( N_{1}-i-1\right) !\left( N_{2}-i\right) !}, \\
b &=&\sum\limits_{i=0}^{\min \left( N_{1}-1,N_{2}-1\right) }\frac{%
A_{n}^{2}N_{1}!N_{2}!\left\vert \zeta \left( t\right) \right\vert ^{2i+1}}{%
i!\left( i+1\right) !\left( N_{1}-i-1\right) !\left( N_{2}-i-1\right) !}, \\
c &=&\sum\limits_{i=0}^{\min \left( N_{1},N_{2}-1\right) }\frac{%
A_{n}^{2}N_{1}!N_{2}!\left\vert \zeta \left( t\right) \right\vert ^{2i}}{%
i!i!\left( N_{1}-i\right) !\left( N_{2}-i-1\right) !}.
\end{eqnarray*}

The relative phase $\varphi _{c}$ is determined by $e^{i\varphi _{c}}=\zeta
/\left\vert \zeta \right\vert $. The coefficient $b$ shows directly the
degree of quantum coherence between two initially independent condensates.
The case $b/a<<1$ means a fragmented state. For $b/a$ near $1$, there is
perfect quantum coherence between two initially independent condensates. One
can prove that there is significant quantum coherence between two
condensates for $N_{1}\left\vert \zeta \right\vert >1$ and $N_{2}\left\vert
\zeta \right\vert >1$. In addition, $b/a$ can be approximated well as $1$
when $N_{1}\left\vert \zeta \right\vert >>1$ and $N_{2}\left\vert \zeta
\right\vert >>1$. A detailed analysis about the nonorthogonality on the
quantum coherence can be found in Refs. \cite{Xiong}.

Without the loss of generality, we consider the one-dimensional case and use
the units where $\hbar =2m=1$. The initial wave functions of two condensates
are assumed as $\phi _{1}\left( x,t=0\right) =\exp \left( -\left( x+s\right)
^{2}/2\Delta ^{2}\right) /\pi ^{1/4}\Delta ^{1/2}$ and $\phi _{2}\left(
x,t=0\right) =\exp \left( -\left( x-s\right) ^{2}/2\Delta ^{2}\right) /\pi
^{1/4}\Delta ^{1/2}$. In our numerical calculations, $\Delta =1$ and $s=6$.
This distance separation makes $\zeta $ between two initial condensates
smaller than $10^{-15}$. Based on Eq. (\ref{zeta}), it is not difficult to
prove that for uniform imaginary component in the complex potential, $\zeta
\left( t\right) $ is always zero for two initially orthogonal wave
functions. To make $\zeta \left( t\right) $ nonzero, nonuniform imaginary
component is necessary. A natural choice is the following imaginary periodic
potential \cite{Oberthaler,Keller,Oberthaler1,Bernet}%
\begin{equation}
V=iV_{0}\sin ^{2}\left( \frac{2\pi x}{d}\right) .
\end{equation}

We calculate numerically the evolution of the system in this imaginary
periodic potential. For the dynamical process of this initial condition, one
can prove rigorously that the relative phase $\varphi _{c}$ is always zero.
In Fig. 2, we give $Log\left( \left\vert \zeta \right\vert \right) $ for
different $d$ and $V_{0}$ at $t=5$. Because the loss of particles has been
considered in the evolution of $\phi _{1}$ and $\phi _{2}$, the quantum
coherence can be determined through the values of $N_{1}\left\vert \zeta
\right\vert $ and $N_{2}\left\vert \zeta \right\vert $ with $N_{1}$ and $%
N_{2}$ being the initial particle numbers. Although $\left\vert \zeta
\right\vert <<1$, the conditions $N_{1}\left\vert \zeta \right\vert >>1$ and
$N_{2}\left\vert \zeta \right\vert >>1$ can be easily satisfied for a wide
range of parameter $d$. Another unique behavior lies in that $\left\vert
\zeta \right\vert $ changes significantly with different $d$.
Experimentally, two equally polarized laser beams intersecting at an angle $%
\theta $ can be used to create the complex periodic potential with $%
d=\lambda /\sin \left( \theta /2\right) $ ($\lambda $ is the laser
wavelength). The inset in Fig. 2 gives $Log\left( \left\vert \zeta
\right\vert \right) $ for $d=5$ and different $V_{0}$. The exponential decay
of $\eta _{1}$ with increasing $V_{0}$ is also shown in the inset.

\begin{figure}[tbp]
\includegraphics[width=0.75\linewidth,angle=270]{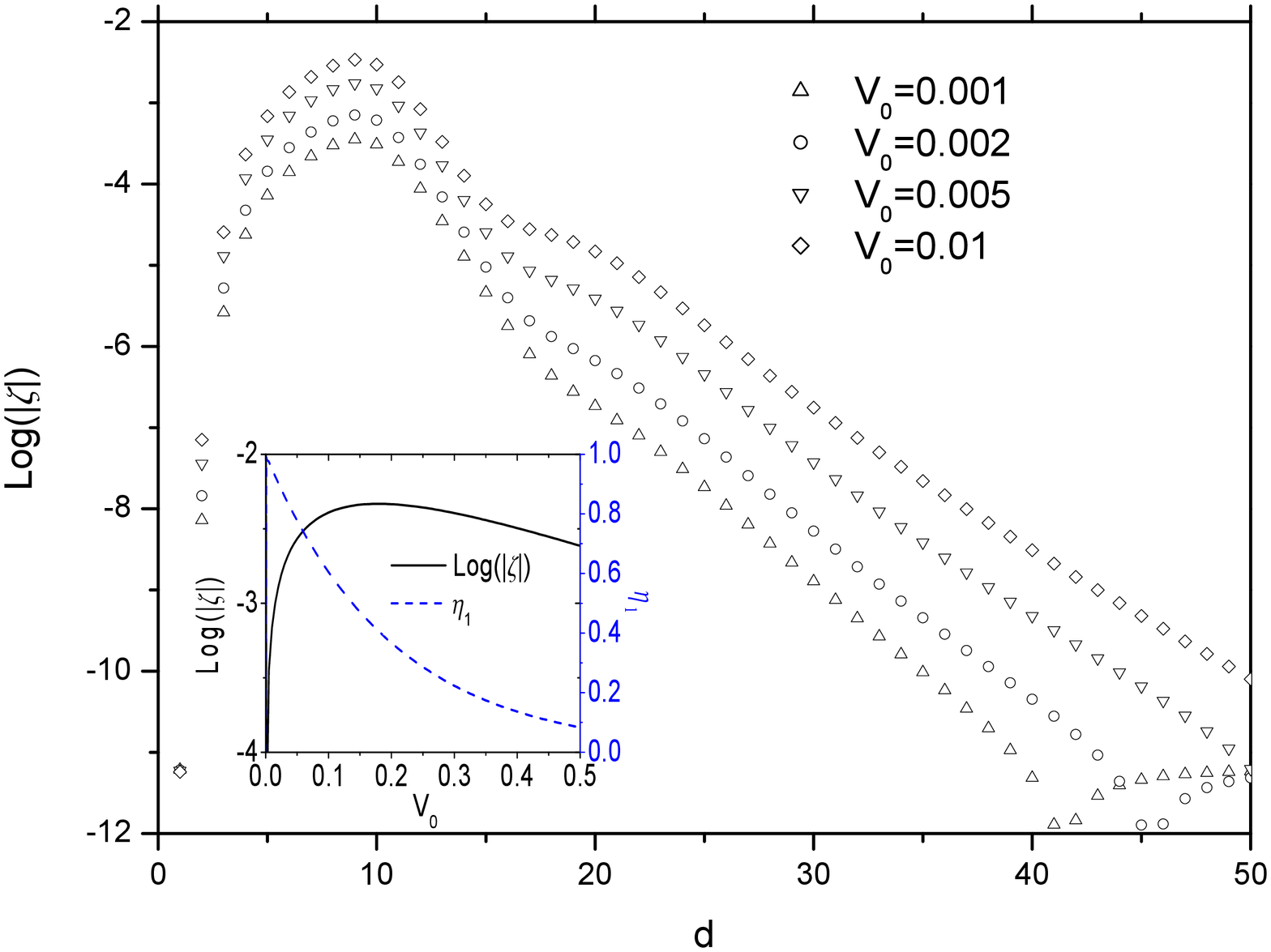}
\caption{Fig. a gives $Log\left( \left\vert \protect\zeta \right\vert
\right) $ for different $d$ and $V_{0}$ at $t=5$. The inset gives $Log\left(
\left\vert \protect\zeta \right\vert \right) $ and $\protect\eta _{1}$ for $%
d=5$ and different $V_{0}$. The units with $\hbar =2m=1$ are adopted.}
\end{figure}

For $\Delta =1$, $s=6$, $d=5$ and $V_{0}=0.001$, we give in Fig. 3a the
evolution of $\eta _{1}\left( t\right) $ and $\left\vert \zeta \right\vert $
based on the numerical calculations of Eq. (\ref{phi1phi2}). Because this
value of $V_{0}$ is extremely small, the decay of $\eta _{1}$ is also very
small. At $t=5$, $\eta _{1}=0.995$ which agrees well with the approximate
analytical expression $\eta _{1}=e^{-V_{0}t}$. This means that the loss of
particles can be omitted, and thus the extra heating effect due to this
imaginary periodic potential can be safely omitted in the whole dynamical
process. As shown in Fig. 3a, we find that $\left\vert \zeta \right\vert <<1$%
. However, for this numerical result of $\left\vert \zeta \right\vert $, the
conditions ($N_{1}\left\vert \zeta \right\vert >>1$ and $N_{2}\left\vert
\zeta \right\vert >>1$) of ideal quantum coherence can be easily satisfied.
In Fig. 3b, the evolution of $b/a$ for $N_{1}=N_{2}=10^{5}$ is shown. We see
that for $t>2.5$, there is very good quantum coherence between two initially
independent condensates. This quantum coherence originates from the nonzero $%
\left\vert \zeta \right\vert $ and exchange symmetry of identical bosons.

\begin{figure}[tbp]
\includegraphics[width=0.75\linewidth,angle=270]{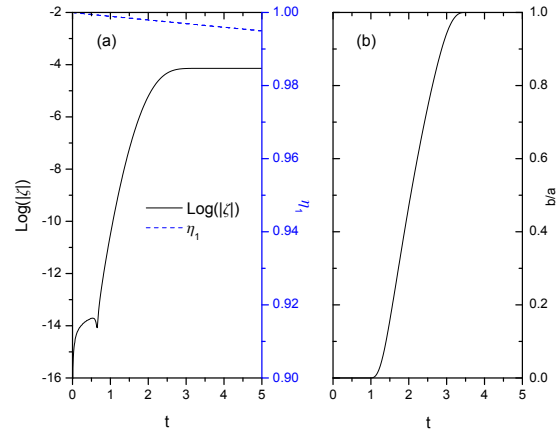}
\caption{Fig. a gives the time evolution of $Log\left( \left\vert \protect%
\zeta \right\vert \right) $ and $\protect\eta _{1}$. Fig. b gives the time
evolution of $b/a$ for initial particle numbers $N_{1}=N_{2}=10^{5}$. It is
shown that there is very good quantum coherence for $t>2.5$, due to the
presence of imaginary periodic potential. The units with $\hbar =2m=1$ are
adopted.}
\end{figure}

In Fig. 4a, we give further the evolution of $n\left( x,t\right) /\left(
N_{1}+N_{2}\right) $ for $N_{1}=N_{2}=10^{5}$. Other parameters are the same
as Fig. 3. Because of the quantum coherence between two condensates, obvious
interference fringes are shown in the density expectation value, which is
quite different from that of Fig. 4b with $V_{0}=0$. Because $V_{0}$ is much
smaller than the kinetic energy $1/2\Delta ^{2}$ of an atom, our numerical
calculations show that the imaginary periodic potential will not play
important role in the shape of $\left\vert \phi _{1}\left( x,t\right)
\right\vert ^{2}$ or $\left\vert \phi _{2}\left( x,t\right) \right\vert ^{2}$%
. This is also the reason why $\left\vert \zeta \right\vert <<1$.

\begin{figure}[tbp]
\includegraphics[width=0.75\linewidth,angle=270]{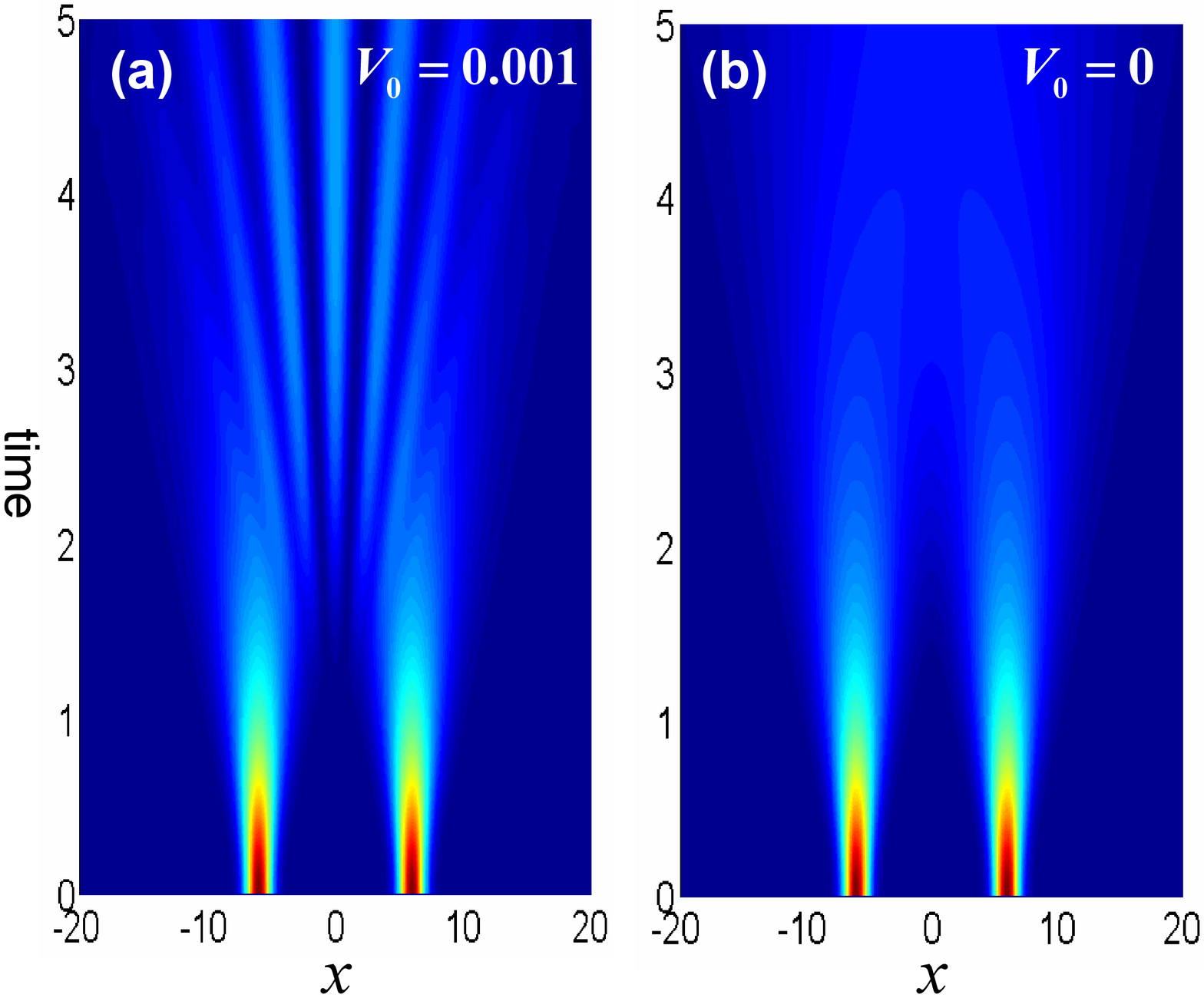}
\caption{Fig. a gives the evolution of $n\left( x,t\right) /\left(
N_{1}+N_{2}\right) $ when the nonorthogonality due to imaginary periodic
potential ($V_{0}=0.001$) and exchange symmetry of identical bosons are both
considered. Fig. b gives the evolution of $n\left( x,t\right) /\left(
N_{1}+N_{2}\right) $ without imaginary potential ($V_{0}=0$). The units with
$\hbar =2m=1$ are adopted.}
\end{figure}

Without the imaginary periodic potential, one should note that after the
overlapping between two initially independent condensates, there would be
still interference fringes with completely random relative phase. This is
due to the well-known measurement-induced interference mechanism \cite%
{Measure}. In a single measurement, there will be clear interference fringes
in the density distribution. By averaging the density distribution of a
large number of measurements, however, there are no more interference
fringes. This is significantly different from the situation that there is
already quantum coherence between two condensates with time evolution, as
shown in Fig. 4b. In this situation, we expect that there is no random shift
in the interference fringes for different experiments with the same initial
conditions. This provides a method to test the imaginary potential induced
quantum coherence in a definite way.

In summary, the widely studied complex potentials for single-body problem
are developed to many-body bosonic system. As a first step toward this new
regime, we consider the role of complex potentials in the quantum coherence
establishment process for two initially independent BECs. Besides the
measure-induced interference mechanism \cite{Measure} and
interaction-induced quantum coherence \cite{Xiong,Ceder,Reinhardt,Danmai}, a
new coherence mechanism---imaginary potential induced quantum coherence is
predicted theoretically. To experimentally test this imaginary potential
induced quantum coherence, Feshbach resonance \cite{Fechbach}\ can be used
to rule out the interaction-induced quantum coherence by tuning $g=0$. Even
without complex potentials induced by the interaction between atoms and
laser, there are various one-, two-, and three-body losses. It is possible
that these losses can be described by a random imaginary potential. Our
numerical calculations show that the inclusion of appropriate random
imaginary potential can also induce the quantum coherence between initially
independent BECs. This has potential application for the formation of a
single condensate in the nonadiabatical evaporative cooling process, where
initially a series of independent subcondensates will form. Considering the
wide existence of random or regular complex potentials, the imaginary
potential induced quantum coherence may have potential applications in
diverse areas of physics where identical particle effect exists.

\begin{acknowledgments}
This work was supported by NSFC under Grant Nos. 10875165, 10634060,
and NKBRSF of China under Grant No. 2006CB921406.
\end{acknowledgments}

\end{document}